\documentclass{tcibook}
\usepackage{fancyhea}
\usepackage{work}
\usepackage{epsfig}

\usepackage{relsize}

\def\beq{\begin{equation}}
\def\eeq#1{\label{#1}\end{equation}}
\def\eeqn{\end{equation}}

\def\beqa{\begin{eqnarray}}
\def\eeqa#1{\label{#1}\end{eqnarray}}
\def\eeqan{\end{eqnarray}}

\let\bar=\overbar

\def\Dslash{\ensuremath{\not{\hbox{\kern-4pt $D$}}}\xspace}
\def\dslash{\not{\hbox{\kern-2pt $\del$}}}

\def\msb{{\bar{\ssstyle M \kern -1pt S}}}

\RequirePackage{xspace}

\usepackage{relsize}
\def\babar{\mbox{\slshape B\kern-0.1em{\smaller A}\kern-0.1em
    B\kern-0.1em{\smaller A\kern-0.2em R}}}

\def\Kbar  {\kern 0.2em\overline{\kern -0.2em K}{}\xspace}

\def\Kzb   {\ensuremath{\Kbar^0}\xspace}
\def\KzKzb {\ensuremath{K^0 \kern -0.16em \Kzb}\xspace}

\def\Dbar  {\kern 0.2em\overline{\kern -0.2em D}{}\xspace}

\def\Dzb   {\ensuremath{\Dbar^0}\xspace}
\def\DzDzb {\ensuremath{D^0 {\kern -0.16em \Dzb}}\xspace}

\def\Bbar  {\kern 0.18em\overline{\kern -0.18em B}{}\xspace}

\def\Bzb   {\ensuremath{\Bbar^0}\xspace}

\def\BzBzb {\ensuremath{B^0 {\kern -0.16em \Bzb}}\xspace}

\mathchardef\Upsilon="7107
\def\Y#1S{\ensuremath{\Upsilon{(#1S)}}\xspace}

\mathchardef\Deltares="7101
\mathchardef\Xi="7104
\mathchardef\Lambda="7103
\mathchardef\Sigma="7106
\mathchardef\Omega="710A
\def\Deltabar   {\kern 0.25em\overline{\kern -0.25em \Deltares}{}\xspace}
\def\Lbar {\kern 0.2em\overline{\kern -0.2em\Lambda\kern 0.05em}\kern-0.05em{}\xspace}
\def\Sigbar{\kern 0.2em\overline{\kern -0.2em \Sigma}{}\xspace}
\def\Xibar{\kern 0.2em\overline{\kern -0.2em \Xi}{}\xspace}
\def\Obar{\kern 0.2em\overline{\kern -0.2em \Omega}{}\xspace}
\def\Nbar{\kern 0.2em\overline{\kern -0.2em N}{}\xspace}
\def\Xb{\kern 0.2em\overline{\kern -0.2em X}{}}

\newcommand{\tev}{\ensuremath{\mathrm{Te\kern -0.1em V}}\xspace}
\newcommand{\gev}{\ensuremath{\mathrm{Ge\kern -0.1em V}}\xspace}
\newcommand{\mev}{\ensuremath{\mathrm{Me\kern -0.1em V}}\xspace}
\newcommand{\kev}{\ensuremath{\mathrm{ke\kern -0.1em V}}\xspace}
\newcommand{\ev}{\ensuremath{\mathrm{e\kern -0.1em V}}\xspace}
\newcommand{\gevc}{\ensuremath{{\mathrm{Ge\kern -0.1em V\!/}c}}\xspace}
\newcommand{\mevc}{\ensuremath{{\mathrm{Me\kern -0.1em V\!/}c}}\xspace}
\newcommand{\gevcc}{\ensuremath{{\mathrm{Ge\kern -0.1em V\!/}c^2}}\xspace}
\newcommand{\mevcc}{\ensuremath{{\mathrm{Me\kern -0.1em V\!/}c^2}}\xspace}

\def\mus  {\ensuremath{\rm \,\mus}\xspace}

\def\mus        {\ensuremath{\,\mu{\rm s}}\xspace}    
\def\gsim{{~\raise.15em\hbox{$>$}\kern-.85em
          \lower.35em\hbox{$\sim$}~}\xspace}
\def\lsim{{~\raise.15em\hbox{$<$}\kern-.85em
          \lower.35em\hbox{$\sim$}~}\xspace}

\def\to                 {\ensuremath{\rightarrow}\xspace}

\def\pep2{PEP-II}
\def\BF{$B$ Factory}
\def\abf {asymmetric \BF}

\xspace

\def\jetset74   {\mbox{\tt Jetset \hspace{-0.5em}7.\hspace{-0.2em}4}}

\begin{document}

\def\bibname{References}
\bibliographystyle{plain}

\raggedbottom

\pagenumbering{roman}

\parindent=0pt
\parskip=8pt
\setlength{\evensidemargin}{0pt}
\setlength{\oddsidemargin}{0pt}
\setlength{\marginparsep}{0.0in}
\setlength{\marginparwidth}{0.0in}
\marginparpush=0pt


\pagenumbering{arabic}

\renewcommand{\chapname}{chap:intro_}
\renewcommand{\chapterdir}{.}
\renewcommand{\arraystretch}{1.25}
\addtolength{\arraycolsep}{-3pt}

\renewcommand{\theequation}{\arabic{equation}}
\renewcommand{\thefigure}{\arabic{figure}}
\renewcommand{\thetable}{\arabic{table}}
\centerline{\bf \Large LEARNING $\gamma$ FROM $B \to K \pi$ DECAYS
\footnote{Contribution to the Proceedings of the Workshop on the Discovery
Potential of an Asymmetric B Factory at $10^{36}$ Luminosity, SLAC, Stanford,
California, 2003}}
\vskip -1.3in
\rightline{CLNS 03/1852}
\rightline{TECHNION-PH-2003-41}
\rightline{hep-ph/0311280}
\vskip 1in

\centerline{Michael Gronau
\footnote{Physics Department, Technion -- Israel Institute of Technology,
Haifa 3200, Israel} and
Jonathan L. Rosner
\footnote {Laboratory of Elementary Particle Physics, Cornell University,
Ithaca, NY 14850, on leave from Enrico Fermi Institute and Department of
Physics, University of Chicago, 5640 S. Ellis Avenue, Chicago, IL 60637}}
\bigskip

Current information on $\gamma = {\rm Arg}(V^*_{ub})$
from other CKM constraints is still in need of improvement, with
$39^\circ < \gamma < 80^\circ$ at 95\% c.l. \cite{Hocker:2001xe}. 
Direct probes of $\gamma$ can tighten these bounds,
possibly indicating new physics effects in case that an 
inconsistency with this range is observed.
In order to study $\gamma$ directly in charmless two-body $B$ decays,
which involve a $b$ to $u$ transition, 
one must generally separate strong and weak phases from one another.
We describe
several cases of $B \to K \pi$ decays in which progress in this work has
been accomplished, and what improvements lie ahead.  Some additional
details are noted in earlier reviews \cite{Gronau:2003cq,Rosner:2003bq,%
Rosner:2003} and in Refs.\ \cite{Gronau:2003kj} and \cite{Gronau:2003kx}.

A great deal of information can be obtained from $B \to K \pi$ decay rates
averaged over CP, supplemented with measurements of direct CP asymmetries.
One probes in this manner tree-penguin interference in various processes.
The data which are used in these analyses are summarized in Table \ref{tab:kpi}
\cite{Chiang:2003pp}. The $B^+$ to $B^0$ lifetime ratio is taken to be
$\tau_+/\tau_0 = 1.078 \pm 0.013$, based on $\tau_+ = 1.653 \pm 0.014$ ps and
$\tau_0 = 1.534 \pm 0.013$ ps \cite{LEPBOSC:2003}.  Table \ref{tab:kpi} also
contains  contributions to the four $B \to K\pi$ decay processes of penguin
($P'$), electroweak penguin ($P'_{\rm EW}$), tree ($T'$) and color-suppressed
tree ($C'$) amplitudes. These contributions are
hierarchical and can be classified using flavor symmetries
\cite{Zeppenfeld:1980ex,Savage:ub,Gronau:1994rj,Gronau:1995hn}.
Smaller contributions, from color-suppressed electroweak penguin amplitudes,
annihilation and exchange amplitudes, are not shown in Table \ref{tab:kpi}.
All four $B\to K\pi$ decays are dominated by penguin amplitudes, which are
related to each other by isospin. Tree amplitudes $T' + C'$ and electroweak
penguin amplitudes $P'_{\rm EW}$  are subdominant and can be related to each
other by flavor SU(3) \cite{Neubert:1998pt}.  SU(3) breaking in tree amplitudes
is introduced assuming factorization.

\begin{table}[h]
\caption{Branching ratios and CP asymmetries for $B \to K \pi$ decays
\cite{Chiang:2003pp}. \label{tab:kpi}}
\begin{center}
\begin{tabular}{c c c c} \hline
Decay mode & Amplitude & ${\cal B}$ (units of $10^{-6}$) & $A_{CP}$ \\ \hline
$B^+ \to K^0 \pi^+$ & $P'$ & $21.78 \pm 1.40$ & $0.016 \pm 0.057$ \\
$B^+ \to K^+ \pi^0$ & $-(P'+P'_{\rm EW}+T'+C')/\sqrt{2}$ & $12.53 \pm 1.04$ &
 $0.00 \pm 0.12$ \\
$B^0 \to K^+ \pi^-$ & $-(P'+T')$ & $18.16 \pm 0.79$ & $-0.095 \pm 0.029$ \\
$B^0 \to K^0 \pi^0$ & $(P'-P'_{\rm EW}-C')/\sqrt{2}$ & $11.68 \pm 1.42$
 & $0.03 \pm 0.37$ \\
\hline
\end{tabular}
\end{center}
\end{table}

Several comparisons between pairs of processes can be made:
\begin{itemize}
\item $B^0 \to K^+ \pi^-~(P'+T'$) vs.\ $B^+ \to K^0 \pi^+~(P'$) 
\cite{Gronau:2003kj,Fleischer:1997um,Gronau:1997an,Gronau:2001cj};
\item $B^+ \to K^+ \pi^0~(P'+P'_{\rm EW}+T'+C')$ vs.\ $B^+ \to K^0 \pi^+~(P')$
\cite{Gronau:2003kj,Neubert:1998pt,Neubert:1998jq,Neubert:1998re};
\item $B^0 \to K^0 \pi^0$ vs.\ other modes \cite{Gronau:2003kj,Buras:1998rb,
Buras:2000gc,Beneke:2001ev,Beneke:2002jn,Beneke:2003zv}.
\end{itemize}
We give the example of $B^0 \to K^+ \pi^-$
in detail.  The tree amplitude for this process is $T' \sim V_{us}V^*_{ub}$,
with weak phase $\gamma$, while the penguin amplitude is $P' \sim V_{ts}
V^*_{tb}$ with weak phase $\pi$.  We denote the penguin-tree relative strong
phase by $\delta$ and define $r \equiv |T'/P'|$.  Then we may write
\begin{eqnarray}
A(B^0 \to K^+ \pi^-) & = & |P'|[1 - r e^{i (\gamma + \delta)}]~, \\
A(\bar B^0 \to K^- \pi^+) & = & |P'|[1 - r e^{i (-\gamma + \delta)}]~, \\
A(B^+ \to K^0 \pi^+) & = & A(B^- \to \bar K^0 \pi^-) = - |P'|~.
\end{eqnarray}
In the last two amplitudes we neglect small annihilation contributions with
weak phase $\gamma$, assuming that rescattering effects are
not largely enhanced. A test for this assumption is the absence of a CP
asymmetry in $B^+ \to K^0 \pi^+$, and a U-spin relation between this
process and $B^+ \to \bar K^0 K^+$ \cite{Falk:1998wc}, in which a
corresponding amplitude with weak phase $\gamma$ is expected to be much larger.
One also neglects small color-suppressed electroweak contributions, for which
experimental tests were proposed in \cite{Fleischer:1998bb}.

\begin{figure}
\begin{center}
\includegraphics[width=0.95\textwidth]{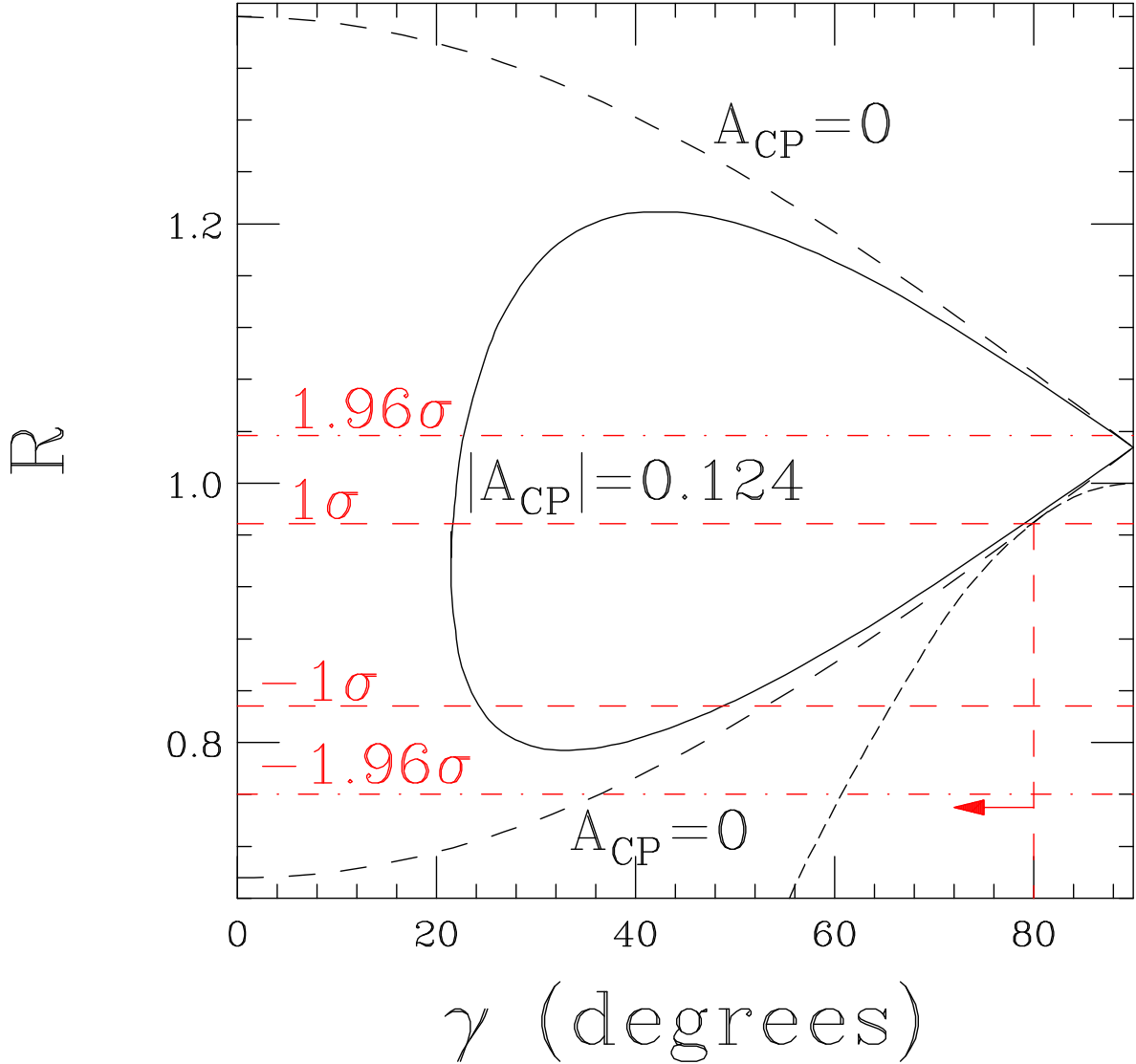}
\caption{Behavior of $R$ for $r = 0.166$ and $A_{CP} = 0$ (dashed curves) or
$|A_{CP}| = 0.124$ (solid curve) as a function of the weak phase $\gamma$.
Horizontal dashed lines denote $\pm 1 \sigma$ experimental limits on $R$,
while dot-dashed lines denote $95\%$ c.l. ($\pm 1.96 \sigma$) limits.  The
short-dashed curve denotes the Fleischer-Mannel bound $\sin^2 \gamma \le R$.
The upper branches of the curves correspond to the case $\cos \gamma \cos
\delta < 0$, while the lower branches correspond to $\cos \gamma \cos \delta
>0$.
\label{fig:Racp}}
\end{center}
\end{figure}

One now forms the ratio
\begin{eqnarray}
R & \equiv & \frac{\Gamma(B^0 \to K^+ \pi^-) + \Gamma(\bar B^0 \to K^- \pi^+)}
{2\Gamma(B^+ \to K^0 \pi^+)} \nonumber \\
 & = & 1 - 2 r \cos \gamma \cos \delta + r^2~.
\end{eqnarray}
Fleischer and Mannel \cite{Fleischer:1997um} pointed out that $R \ge \sin^2
\gamma$ for any $r, \delta$ so if $1 > R$ one can get a useful bound.
Moreover, if one uses
\begin{equation}
R A_{CP} (K^+\pi^-)= - 2 r \sin \gamma \sin \delta
\end{equation}
as well and eliminates $\delta$ one can get a more powerful constraint,
illustrated in Fig.\ \ref{fig:Racp}.

We have used $R = 0.898 \pm 0.071$ and $A_{CP} = -0.095 \pm 0.029$
based on recent averages \cite{Chiang:2003pp} of CLEO, BaBar, and
Belle data, and $r = |T'/P'| = 0.142^{+0.024}_{-0.012}$.  In order to
estimate the tree amplitude and the ratio of amplitudes $r$, we have used
factorization in $B^0 \to \pi^- \ell^+ \nu_\ell$ at low $q^2$ \cite{Luo:2003hn}
and $\left|\frac{T'}{T} \right| = \frac{f_K}{f_\pi}\left| \frac{V_{us}}{V_{ud}}
\right| \simeq(1.22)(0.23) = 0.28$.  One could also use processes in which $T$
dominates, such as $B^0 \to \pi^+ \pi^-$ or $B^+ \to \pi^+ \pi^0$, but these
are contaminated by contributions from $P$ and $C$, respectively.
The $1\sigma$ allowed region lies between the curves $A_{CP} = 0$ and
$|A_{CP}| = 0.124$.  The most conservative upper bound on $\gamma$
arises for the smallest value of $|A_{CP}|$ and the largest value of $r$,
while the most conservative lower bound would correspond to the largest
$|A_{CP}|$ and the smallest $r$. Currently no such lower bound is obtained
at a $1\sigma$ level. At this level one has $R < 1$, leading to an upper
bound $\gamma < 80^\circ$.

We note that for the current average value of $R$ the $1\sigma$ upper bound,
$\gamma < 80^\circ$, happens to coincide with that of
Ref.~\cite{Fleischer:1997um}. This bound does not depend much on the value of
$r$, for which we assumed factorization of $T$ in order to introduce SU(3)
breaking.  The upper bound on $\gamma$ varies only slightly, $\gamma <
78^\circ - 80^\circ$, for a wide range of values $r = 0.1 - 0.3$.  On the other
hand, a potential lower bound on $\gamma$ depends more sensitively on the
value of $r$, and would result if small values of this parameter could be
excluded.  For instance, Fig.\ \ref{fig:Racp} shows that a value $r=0.166$
implies $\gamma > 49^\circ$ at $1\sigma$. Thus, it is crucial to improve our
knowledge of $r$.

\begin{figure}
\begin{center}
\includegraphics[width=0.95\textwidth]{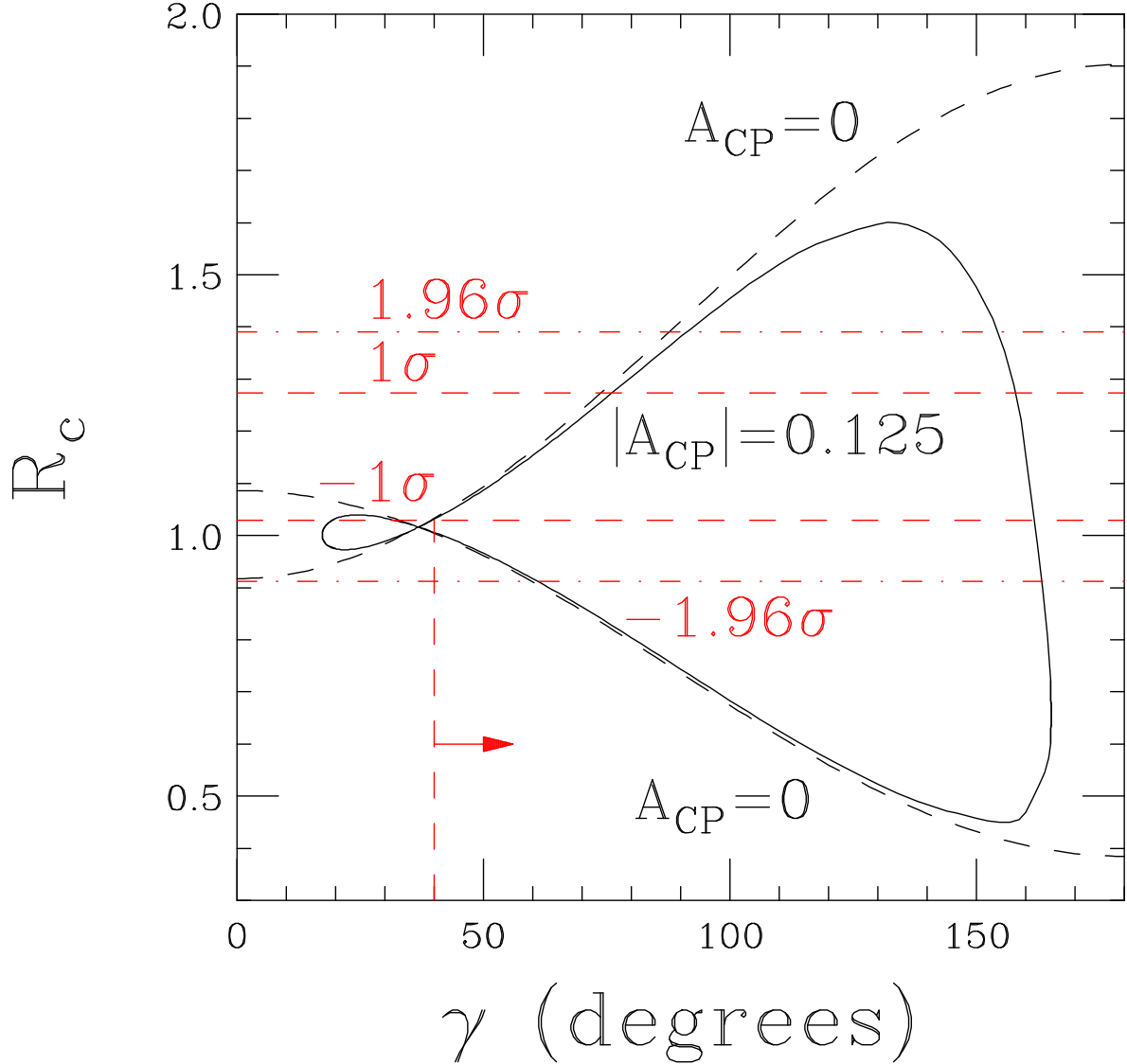}
\caption{Behavior of $R_c$ for $r_c = 0.21$ ($1 \sigma$ upper limit) and
$A_{CP}(K^+ \pi^0) = 0$ (dashed curves) or $|A_{CP}(K^+ \pi^0)|
= 0.125$ (solid curve) as a function of the weak phase $\gamma$.
Horizontal dashed lines denote $\pm 1 \sigma$ experimental limits on $R_c$,
while dotdashed lines denote 95\% c.l. ($ \pm 1.96 \sigma$) limits.  We have
taken $\delta_{EW} = 0.80$ (its $1 \sigma$ upper limit), which
leads to the most conservative bound on $\gamma$.
Upper branches of curves correspond to $\cos \delta_c (\cos \gamma -
\delta_{EW}) < 0$, while lower branches correspond to $\cos \delta_c (\cos
\gamma - \delta_{EW}) >0$.  Here $\delta_c$ is a strong phase.
\label{fig:Rcacp}}
\end{center}
\end{figure}

The process $B^+ \to K^+ \pi^0$ also provides constraints on $\gamma$.  The
deviation of the ratio
\begin{equation}
R_c \equiv \frac{\Gamma(B^+ \to K^+ \pi^0)+\Gamma(B^- \to K^- \pi^0)}
{\Gamma(B^+ \to K^0 \pi^+)} = 1.15 \pm 0.12
\end{equation}
from 1, when combined with $A_{CP} (K^+\pi^0)= 0.00 \pm 0.12$, $r_c = |(T'+C')
/P'| = 0.195\pm 0.016$ and an estimate of the electroweak penguin amplitude
$\delta_{EW} \equiv |P'_{EW}|/ |T'+C'| = 0.65 \pm 0.15$, leads to a $1 \sigma$
lower bound $\gamma > 40^\circ$.  Details of the method may be found in
Refs.\ \cite{Gronau:2003cq,
Rosner:2003bq,Gronau:2003kj,Neubert:1998pt,Neubert:1998jq,Neubert:1998re};
the present bound represents an update of previously quoted values.
The most conservative lower bound on $\gamma$ arises for smallest $A_{CP}$,
largest $r_c$, and largest $|P'_{EW}|$, and is shown in Fig.\ \ref{fig:Rcacp}.
These values of  $r_c$ and $|P'_{EW}|$ would also imply an upper bound,
$\gamma < 77^\circ$, which demonstrates the importance of improving our
knowledge of these two hadronic parameters.

\begin{figure}
\begin{center}
\includegraphics[width=0.95\textwidth]{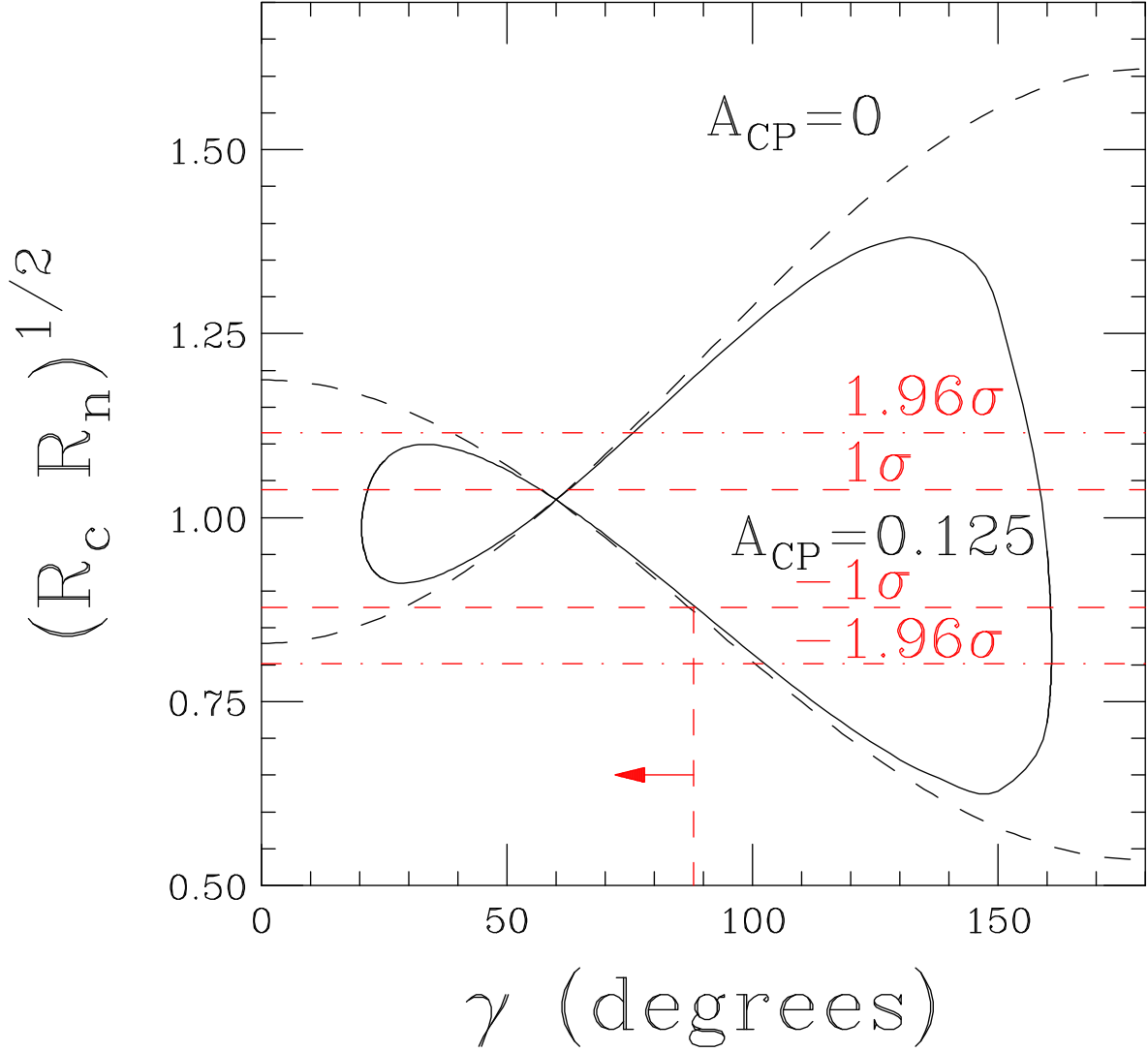}
\caption{Behavior of $(R_c R_n)^{1/2}$ for $r_c = 0.18$ ($1 \sigma$ lower
limit) and $A_{CP}(K^+ \pi^0) = 0$ (dashed curves) or
$|A_{CP}(K^+ \pi^0)| = 0.125$ (solid curve) as a function of the
weak phase $\gamma$. Horizontal dashed lines denote $\pm 1 \sigma$ experimental
limits on $(R_c R_n)^{1/2}$, while dotdashed lines denote 95\% c.l. ($ \pm 1.96
\sigma$) limits.  Upper branches of curves correspond to $\cos \delta_c(\cos
\gamma - \delta_{EW}) < 0$, while lower branches
correspond to $\cos \delta_c(\cos \gamma - \delta_{EW}) > 0$.  Here we have
taken $\delta_{EW} = 0.50$ (its $1 \sigma$ lower limit), which
leads to the most conservative bound on $\gamma$.
\label{fig:Rmean}}
\end{center}
\end{figure}

Another ratio
\begin{equation}
R_n  \equiv  \frac{\Gamma(B^0\to K^+ \pi^-)+\Gamma(\bar B^0\to K^- \pi^+)}
{2 \left [\Gamma(B^0 \to K^0 \pi^0)+\Gamma(\bar B^0 \to \bar K^0 \pi^0)\right]}
= 0.78 \pm 0.10
\end{equation}
involves the decay $B^0 \to K^0 \pi^0$.  This ratio should be equal to $R_c$
since to leading order in $T'/P'$, $C'/P'$, and $P'_{EW}/P'$ one has
\begin{equation}
\left|\frac{P'+T'} {P'-P'_{\rm EW}-C'} \right|^2 \approx
 \left| \frac{P'+P'_{\rm EW}+T'+C'}{P'} \right|^2~~,
\end{equation}
but the two ratios differ by $2.4 \sigma$.
Possibilities for explaining this apparent discrepancy (see, e.g., Refs.\
\cite{Gronau:2003kj,Grossman:2003lp})
include (1) new physics, e.g., in the EWP amplitude, and (2) an underestimate
of the $\pi^0$ detection efficiency in all experiments, leading to an
overestimate of any branching ratio involving a $\pi^0$.  The latter
possibility can be taken into account by considering the ratio
$(R_n R_c)^{1/2} = 0.96 \pm 0.08$, in which the $\pi^0$ efficiency cancels.
As shown in Fig.\ \ref{fig:Rmean}, this ratio leads only to the conservative
bound $\gamma \le 88^\circ$.  A future discrepancy between $R_c$ and $R_n$ at a
statistically significant level implying new physics effects would clearly
raise questions about the validity of constraints on $\gamma$ obtained from
these quantities.

Recently a time-dependent asymmetry measurement in $B^0(t) \to K_S\pi^0$ was 
reported \cite{SCkpi}
\begin{equation}
S_{\pi K} = 0.48^{+0.38}_{-0.47} \pm 0.11~,~~~~C_{\pi K} = 0.40^{+0.27}_{-0.28}
\pm 0.10~~,
\end{equation}
where $S_{\pi K}$ and $-C_{\pi K}$ are coefficients of $\sin\Delta mt$ and
$\cos\Delta mt$ terms in the asymmetry. In the limit of a pure penguin 
amplitude, $A(B^0 \to K^0\pi^0) = (P' - P'_{\rm EW})/\sqrt{2}$,  one expects 
$S_{\pi K} = \sin 2\beta,  C_{\pi K}=0$. The color-suppressed amplitude, 
$C'$, contributing to this process involves a weak phase $\gamma$. 
Its effect was studied recently \cite{Gronau:2003kx} by relating these two 
amplitudes within flavor SU(3) symmetry to corresponding amplitudes in 
$B^0 \to \pi^0\pi^0$. Correlated deviations from $S_{\pi K} = \sin 2\beta,  
C_{\pi K}=0$, at a level of $0.1-0.2$ in the two asymmetries, were calculated 
and were shown to be sensitive to values of $\gamma$ in the currently 
allowed range. Observing such deviations and probing the value of 
$\gamma$ requires reducing errors in the two asymmetries by about an order 
of magnitude. 

To summarize, promising bounds on $\gamma$ stemming from various 
$B \to K \pi$ decays have been mentioned.  So far all are statistics-limited.  
At $1 \sigma$ we have found

\begin{itemize}
\item $R$ ($K^+ \pi^-$ vs.\ $K^0 \pi^+$) gives $\gamma \le 80^\circ$;

\item $R_c$ ($K^+ \pi^0$ vs.\ $K^0 \pi^+$) gives $\gamma \ge 40^\circ$;

\item $R_n$ ($K^+ \pi^-$ vs.\ $K^0 \pi^0$) should equal $R_c$;
$(R_c R_n)^{1/2}$ gives $\gamma \le 88^\circ$.

\end{itemize}

The future of most such $\gamma$ determinations remains for now in
experimentalists' hands, as one can see from the Figures.  We have noted
(see, e.g., \cite{Gronau:1997an}) that measurements of rate ratios
in $B \to K \pi$ can ultimately pinpoint $\gamma$ to within about $10^\circ$.
The required accuracies in $R$, $R_c$, and $R_n$ to achieve this goal can be
estimated from the Figures.  For example, knowing $(R_c R_n)^{1/2}$ to within
0.05 would pin down $\gamma$ to within $10^\circ$ if this ratio lies in the
most sensitive range of Fig.\ \ref{fig:Rmean}. 
A significant discrepancy between the values of $R_c$ and 
$R_n$ would be evidence for new physics. 

It is difficult to extrapolate the usefulness of $R$, $R_c$, and $R_n$
measurements to very high luminosities without knowing ultimate
limitations associated with systematic errors.  The averages in Table
\ref{tab:kpi} are based on individual measurements in which the statistical
errors exceed the systematic ones by at most a factor of about 2 (in the
case of $B^0 \to K^0 \pi^0$) \cite{Chiang:2003pp}.  For $B^+ \to K^+ \pi^0$ the
statistical and systematic errors are nearly equal.  Thus, the clearest path
to improvements in these measurements is associated with the next factor of
roughly 4 increase in the total data sample.  Thereafter, reductions in
systematic errors must accompany increased statistics in order for these
methods to yield improved accuracies in $\gamma$.

In our study we used the most pessimistic values of the parameters $r$, $r_c$
and $\delta_{EW}$ leading to the weakest bounds on $\gamma$.
The theoretical uncertainties in these parameters can be
further reduced, and the assumption of negligible rescattering can be tested.
This progress will rely on improving branchng ratio measurements for
$B\to K\pi$, $B\to \pi\pi$ and $B^0 \to \pi^- \ell^+ \nu_\ell$, on
an observation of penguin-dominated $B \to K\bar K$ decays, and on various
tests of factorization which imply relations between CP-violating rate
differences \cite{Deshpande:1994ii,Gronau:2000zy}.

A complementary approach to the flavor-SU(3) method is the QCD
factorization formalism of Refs.\ \cite{Beneke:2001ev,Beneke:2002jn,%
Beneke:2003zv}.  It predicts small strong phases (as
found in our analysis) and deals directly with flavor-SU(3) breaking; however,
it involves some unknown form factors and meson wave functions and appears
to underestimate the magnitude of $B \to VP$ penguin amplitudes.
Combining the two approaches seems to be the right way to proceed.

This work was supported in part by the United States Department of Energy
through Grant No.\ DE FG02 90ER40560.


\begin{thebibliography}{99}


\bibitem{Hocker:2001xe}
A.~Hocker, H.~Lacker, S.~Laplace and F.~Le Diberder,
Eur.\ Phys.\ J.\ C {\bf 21}, 225 (2001)
[arXiv:hep-ph/0104062].  For periodic updates see
{\tt http://ckmfitter.in2p3.fr/.}

\bibitem{Gronau:2003cq}
M.~Gronau,
Presented at Flavor Physics and CP Violation 
(FPCP 2003), Paris, France, 3--6 June 2003
[arXiv:hep-ph/0306308]. 

\bibitem{Rosner:2003bq}
J.~L.~Rosner,
AIP Conf.\ Proc.\  {\bf 689}, 150 (2003)
[arXiv:hep-ph/0306284].

\bibitem{Rosner:2003}
J.~L.~Rosner,
Presented at the 9th International Conference on  B-Physics at Hadron 
Machines (BEAUTY 2003), CMU, Pittsburgh, October 2003
[arXiv:hep-ph/0311170]. 

\bibitem{Gronau:2003kj}
M.~Gronau and J.~L.~Rosner,
Phys.\ Lett.\ B {\bf 572}, 43 (2003)
[arXiv:hep-ph/0307095].

\bibitem{Gronau:2003kx}
M.~Gronau, Y.~Grossman and J.~L.~Rosner,
to be published in Phys. Lett. B 
[arXiv:hep-ph/0310020].

\bibitem{Chiang:2003pp}
C.-W. Chiang, M. Gronau, J. L. Rosner, and D. A. Suprun, 
2003, manuscript in preparation; see also Heavy Flavor
Averaging Group, Lepton-Photon 2003 branching ratios and
CP asymmetries, at {\tt http://www.slac.stanford.edu/xorg/hfag/rare/}.

\bibitem{LEPBOSC:2003}
LEP $B$ Oscillation Working Group,
results for the summer 2003 conferences,
{\tt http://lepbosc.web.cern.ch/LEPBOSC/}.

\bibitem{Zeppenfeld:1980ex}
D.~Zeppenfeld,
Z.\ Phys.\ C {\bf 8}, 77 (1981).

\bibitem{Savage:ub}
M.~J.~Savage and M.~B.~Wise,
Phys.\ Rev.\ D {\bf 39}, 3346 (1989)
[Erratum-ibid.\ D {\bf 40}, 3127 (1989)].

\bibitem{Gronau:1994rj}
M.~Gronau, O.~F.~Hernandez, D.~London and J.~L.~Rosner,
Phys.\ Rev.\ D {\bf 50}, 4529 (1994)
[arXiv:hep-ph/9404283].

\bibitem{Gronau:1995hn}
M.~Gronau, O.~F.~Hernandez, D.~London and J.~L.~Rosner,
Phys.\ Rev.\ D {\bf 52}, 6374 (1995)
[arXiv:hep-ph/9504327].

\bibitem{Neubert:1998pt}
M.~Neubert and J.~L.~Rosner,
Phys.\ Lett.\ B {\bf 441}, 403 (1998)
[arXiv:hep-ph/9808493].

\bibitem{Fleischer:1997um}
R.~Fleischer and T.~Mannel,
Phys.\ Rev.\ D {\bf 57}, 2752 (1998)
[arXiv:hep-ph/9704423].

\bibitem{Gronau:1997an}
M.~Gronau and J.~L.~Rosner,
Phys.\ Rev.\ D {\bf 57}, 6843 (1998)
[arXiv:hep-ph/9711246].

\bibitem{Gronau:2001cj}
M.~Gronau and J.~L.~Rosner,
Phys.\ Rev.\ D {\bf 65}, 013004 (2002)
[Erratum-ibid.\ D {\bf 65}, 079901 (2002)]
[arXiv:hep-ph/0109238].

\bibitem{Neubert:1998jq}
M.~Neubert and J.~L.~Rosner,
Phys.\ Rev.\ Lett.\  {\bf 81}, 5076 (1998)
[arXiv:hep-ph/9809311].

\bibitem{Neubert:1998re}
M.~Neubert,
JHEP {\bf 9902}, 014 (1999)
[arXiv:hep-ph/9812396].

\bibitem{Buras:1998rb}
A.~J.~Buras and R.~Fleischer,
Eur.\ Phys.\ J.\ C {\bf 11}, 93 (1999)
[arXiv:hep-ph/9810260].

\bibitem{Buras:2000gc}
A.~J.~Buras and R.~Fleischer,
Eur.\ Phys.\ J.\ C {\bf 16}, 97 (2000)
[arXiv:hep-ph/0003323].

\bibitem{Beneke:2001ev}
M.~Beneke, G.~Buchalla, M.~Neubert and C.~T.~Sachrajda,
Nucl.\ Phys.\ B {\bf 606}, 245 (2001)
[arXiv:hep-ph/0104110].

\bibitem{Beneke:2002jn}
M.~Beneke and M.~Neubert,
Nucl.\ Phys.\ B {\bf 651}, 225 (2003)
[arXiv:hep-ph/0210085].

\bibitem{Beneke:2003zv}
M.~Beneke and M.~Neubert,
arXiv:hep-ph/0308039.

\bibitem{Falk:1998wc}
A.~F.~Falk, A.~L.~Kagan, Y.~Nir and A.~A.~Petrov,
Phys.\ Rev.\ D {\bf 57}, 4290 (1998)
[arXiv:hep-ph/9712225].

\bibitem{Fleischer:1998bb}
R.~Fleischer,
Eur.\ Phys.\ J.\ C {\bf 6}, 451 (1999)
[arXiv:hep-ph/9802433].

\bibitem{Luo:2003hn}
Z.~Luo and J.~L.~Rosner,
Phys.\ Rev.\ D {\bf 68}, 074010 (2003)
[arXiv:hep-ph/0305262].

\bibitem{Grossman:2003lp}
Y.~Grossman,
Presented at LP03, XXI International Symposium on Lepton and Photon 
Interactions at High Energies, Fermilab, 2003 [hep-ph/0310229].

\bibitem{SCkpi} BaBar Collaboration, A. Farbin {\it et al.}, PLOT--0053, 
contribution to LP03,
Ref.~\cite{Grossman:2003lp}; 
{\tt https://oraweb.slac.stanford.edu:8080/pls/slacquery/babar\_documents.startup}.

\bibitem{Deshpande:1994ii}
N.~G.~Deshpande and X.~G.~He,
Phys.\ Rev.\ Lett.\  {\bf 75}, 1703 (1995)
[arXiv:hep-ph/9412393].

\bibitem{Gronau:2000zy}
M.~Gronau,
Phys.\ Lett.\ B {\bf 492}, 297 (2000)
[arXiv:hep-ph/0008292].

\end{thebibliography}
\end{document}